# Considerations for Cloud Security Operations

**James J. Cusick, PMP**
Chief Security Officer & Director IT Operations
Wolters Kluwer, CT Corporation, New York, NY
j.cusick@computer.org

**Abstract—Information Security in Cloud Computing environments is explored. Cloud Computing is presented, security needs are discussed, and mitigation approaches are listed.**

## I. Introduction

The promise of cloud computing offers scalability, reduced costs, flexibility, interoperability, transportability, self-configuration, and more. Unfortunately, all these attributes also introduce new security risks for the users of such an environment. It is important to know what is meant by cloud computing, what the vulnerabilities are in such environments, and what countermeasures are available to protect data that might be used in a cloud environment and to facilitate use of such promising and not prevalent technologies. This paper will touch on some of these topics in a summary fashion for those readers interested to get a quick understanding of the scope of the problem.

## II. Background

The author is the CSO (Chief Security Officer) of CT Corporation (CT) which is a division of Wolters Kluwer (WK). WK is a Netherlands-based international publisher and digital information services provider with operations around the world. CT is engaged in corporate legal services in support of many types of companies. The systems supported include public-facing Web-based applications and internally used ERP (Enterprise Resource Planning) systems. Major technical vendors manage network services, private hosting, and cloud services for CT.

The CT IT operations team manages a large scale computing environment using an ITIL model [1]. Over the last several years CT has also developed a robust security program managed from within this operations team [2]. This program focuses on protecting company and customer information assets, reducing risk, educating users, and providing a CSIRT (Computer Security Incident Response Team) function throughout CT. This has also included establishing an Executive Governance process and the creation of a Secure SDLC (Software Development Lifecycle) which integrated OWASP coding practices, automated vulnerability scans, and more. This work was done under the guidance of a detailed security control objective framework derived from ISO 27001 standards and related sources.

Of late a significant concern facing the CT security program has been the aspects of security as they relate to cloud computing. CT currently takes advantage of a private cloud, numerous SaaS applications, and social media in Community Cloud environments (see Figure 1). CT is also engaged in migration of some applications that were designed as cloud-ready to move from a private cloud to a public cloud.

These business and technical needs drive both security and operational questions to the forefront. As a CSO it is critical to understand the implications of utilizing these environments and technologies and what judgements are required to do so.

## III. The Cloud Defined

The NIST (National Institute of Standards and Technology) defines cloud computing as such:

> *Cloud computing is a model for enabling ubiquitous, convenient, on-demand network access to a shared pool of configurable computing resources (e.g., networks, servers, storage, applications, and services) that can be rapidly provisioned and released with minimal management effort or service provider interaction. This cloud model is composed of five essential* *characteristics, three service models, and four deployment models [3].*

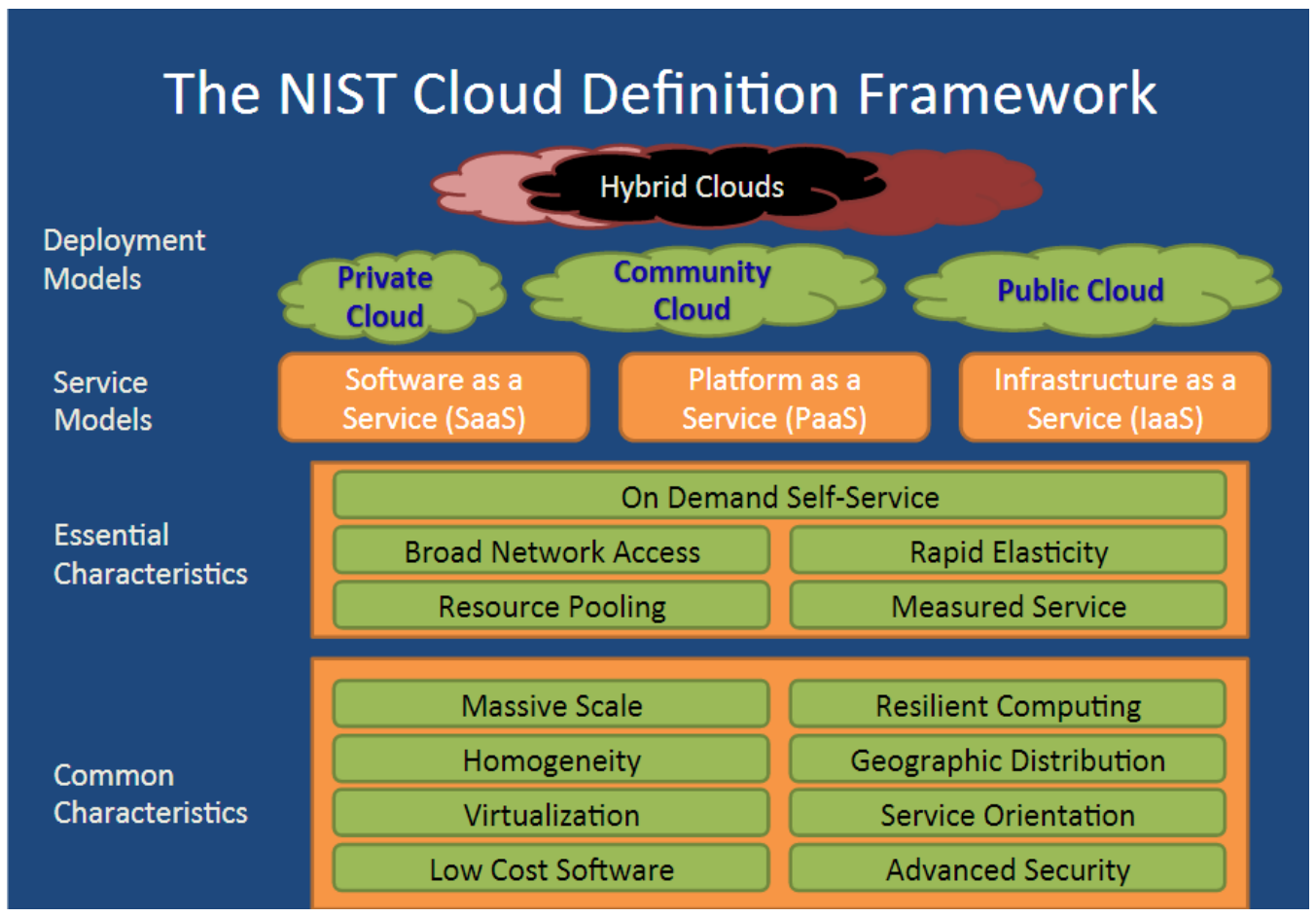

*Figure 1 – NIST Cloud Definition Framework [4]*

The essential characteristics of a cloud computing environment include:

1. On-demand self-service
2. Broad network access
3. Resource pooling
4. Rapid elasticity
5. Measured service

Naturally, some of these very characteristics open the possibility for attack, abuse, or other security issues. For example, with self-service, if not properly bounded administrative controls can leak over to other environments. This is the same for broad network access and resource pooling. These attributes both open the door to a variety of potential security issues.

An interesting characteristic which is *missing* from this NIST cloud model is security. While it is mentioned as a configuration item of community clouds it is not called out in any detail. What this means is that security essentially comes down to the buyer or user of these environments. It is possible that a cloud vendor may have a well secured environment, but the risk of use is primarily placed on the service customer. However, the “NIST Cloud Computing Security Reference Architecture” document (500-299) does provide a detailed model for security management in the cloud.

## IV. Vulnerabilities in the Cloud

For any application, security vulnerabilities may exist. Following the types of security practices which CT has put in place [2] reduces this risk of deploying applications with such vulnerabilities, but it does not eliminate that risk. To prepare for cloud usage it is important to understand what some of these vulnerabilities might be and where they might be more likely, for example, in the use of SaaS applications or the hosting of custom applications. Each application architectures may have vulnerability possibilities: Web Services, SOA, REST, APIs, virtualization, multi-tenant egress to single VM, all present risks [4].

Furthermore, within these architecture types the typical Web application vulnerabilities remain present and must be defended against including:

- SQL injection,
- Cross-site scripting,
- Cross-site request forgery,
- Session hijacking

Identity management is also a major issue (trusted domain management) in cloud environments. Thus, all the operational and security practices needed in a private cloud or traditional hosted environment still apply in the cloud but in new ways.

## V. Cloud Computing Concerns

From a business and operational perspective, the primary concerns in cloud computing security are availability, integrity, and confidentiality [4]. This is not entirely different than in a traditionally hosted environment but achieving these goals requires additional work in cloud computing environments as described in the next section.

From a technical standpoint, Vitti suggests the following list of key concerns within cloud computing [4]:

1. Availability.
2. Access control.
3. Vulnerability management.
4. Patch and configuration management.
5. Countermeasures.
6. Cloud usage and access monitoring.
7. Malicious use of abundant computing resources.

Each of these areas has specific mitigations which can be put in place. In CT’s view, an overriding set of concerns are the binding agreements governing specific types of data to be placed in or processed in a cloud environment. The above list only serves to meet the need of data protection. Without knowing what data is in scope for use within the cloud an effective cloud security model cannot be developed. This is further buttressed by any legal commitments to customers, either implied or explicit when managing their data.

## VI. Managing the Concerns

In exploring the above concerns several standard approaches to mitigating, managing, or reducing risk associated with them stand out. For CT this begins with understanding the field of use. There is much interest in using cloud environments whether for SaaS, hosting, or collaboration with partners and customers. The first question the CSO must understand is what purpose the selected cloud technology is meant to serve.

With that understood the applicable requirements, legal structure, and data management approach begins to emerge.

The next focus point is on the data involved. If sensitive data is involved specific approaches must be taken to manage the data and in particular the legal framework must be understood. Only in cases where customer allow data supporting their services can this be placed in the cloud. This requires careful contractual reviews with the General Counsel and coordination with product management, engineering, and development on the data content to be used [5]. This is also true for any SaaS usage.

The next major step is vendor assessment. For each vendor CT utilizes the Standardized Information Gathering (SIG) Questionnaire from Shared Assessments (The Santa Fe Group). This assessment includes several hundred questions covering business profile, risk, security, human resources, and more. Upon review of the replies, if there are any questions a discussion is held with the vendor to better understand any gaps. Compliance with this framework leads to a strong understanding of any cloud vendor and their operational and security practices and provides a uniform assessment mechanism for auditing purposes.

Finally, the established ITIL based IT operations model [1] with CT follows and its security controls [2] are overlaid onto the cloud vendor. Communications, SLAs, escalation models, incident management, and supporting processes are all reviewed and mapped to those of the cloud provider.

The specific technical approaches which need immediate focus in building out a cloud processing solution include [4]:

- An encryption schema for all communications and storage locations
- Backup plan
- Understanding of the risks of data exposure

To further these management approaches, Souza [6] recommends the development of a cloud brokerage platform to abstract internal systems from cloud environments, developing federated identity schemes, and building data brokerage environments. In terms of priority some of the key practices to have in place include the following [6]:

1. Develop a clear data classification scheme
2. Provide a data ownership model
3. Ensure policy management for all access levels including administration
4. Ensure data lifecycle management including retention and destruction
5. Coordinate logging of all actions

## VII. Customer Needs and Audit Requirements

Some of the major drivers for CT's security program include security best practices, customer requirements, product ideas, and auditing frameworks like SOC (Service Organization Control) reports. As a result the security team must work closely with legal, internal audit, external audit, customers, engineering, operations, and management on realizing defined security objectives and continue to operate and improve the program especially as it relates to utilizing cloud computing [5]. These inputs in turn drive security controls, methods, tools, and practices necessary to meet these requirements. As cloud technologies, SaaS applications, and related technologies have grown in popularity CT's security standards and practices have evolved accordingly along the lines outlined above in this paper. Going forward we expect this evolution to continue and to increase in velocity calling for further sophistication in our approach.

## VIII. Conclusions

Our journey to develop and implement appropriate security methods for managing cloud computing started over a decade ago with our first SaaS support tool. At that time CT also began offering its own SaaS applications. Thus from the beginning of the cloud revolution we have been both a cloud vendor and consumer. For both such situations security plays a key role and will continue to do so.

## IX. Acknowledgements

The CT Corp Security Program which drove the development of CT's Cloud Security understanding was supported by both the leadership team and the expert engineering staff.

## XI. AUTHOR CONTACT

James Cusick, was previously CISO & Director IT Operations, Wolters Kluwer's CT Corporation, New York, NY. He is currently Visiting Associate Researcher at Ritsumeikan University in Kyoto, Japan. Reach James at: j.cusick@computer.org.